\documentclass{elsart}
\usepackage{amssymb}

\usepackage{amsmath}

\input psfig.sty
\input{tcilatex}

\begin{document}

\begin{frontmatter}

\title{Relativistic short-range correlation effects on the pion dynamics in nuclear matter}

\author{L. B. Leinson$^{1}$ and A. P\'{e}rez$^{2}$}

\address{$^{1}$Departamento de Astronom\'{i}a y Astrof\'{i}sica, Universidad de Valencia\\
46100 Burjassot (Valencia), Spain\\
and\\ Institute of Terrestrial Magnetism, Ionosphere and
Radio Wave Propagation
RAS, 142190 Troitsk, Moscow Region, Russia\\
$^{2}$Departamento de F\'{i}sica Te\'{o}rica and IFIC, Universidad de Valencia 
46100 Burjassot (Valencia), Spain \\
E-mail:\\ 
leinson@izmiran.rssi.ru\\
Armando.Perez@uv.es}

\begin{abstract}
Replicated theoretical attempts of relativistic approaches to the pion self-energy in 
nuclear matter yield unphysical pion spectra. We demonstrate the crucial dependence 
of the calculated pion spectra on the correct relativistic accounting for the short-range 
correlation effects on the pion self-energy in the medium. 
To do this, we simulate the short-range interactions by phenomenological 
contact terms in the relativistic Lagrangian density, and derive 
the pion self-energy by carefully taking into account the relativistic kinematics. The 
obtained spectrum for the pion-like excitations in cold nuclear matter shows physically 
meaningful branches, in contrast to those obtained before by different authors by the 
use of simplified relativistic approaches to the short-range correlations.

PACS number(s): 24.10.Cn; 13.75.Cs; 21.65.+f; 25.70.-z 
\end{abstract}

\end{frontmatter}\newpage

\section{Introduction}

The problem of the pion spectrum in nucleon matter has received renewed
interest due to experiments with relativistic heavy-ion collisions.
Considerable efforts from many theoretical groups were made to study the
relativistic dispersion relation for in-medium pions (see e. g. \cite{Mao}
and references therein). Nevertheless, a complete theoretical treatment of
the pion-like excitations in nuclear matter is still not available.
Replicated theoretical attempts of relativistic approaches to the pion
self-energy in nuclear matter yield the pion spectra with two pion-like
branches merging at some point where $d\omega /dk=\infty $ (see e. g. \cite%
{Mao}, \cite{Herb}, \cite{Mornas}). This kind of spectrum implies that the
group velocity of a pion quasi-particle can be larger than that of light,
contrarily to canons of the relativistic theory. The situation might be
understood in view that the relativistic theory of short-range correlations
in nuclear matter remains unsolved, and it forces the authors to use
nonrelativistic analogies, although they have never been checked carefully
(see e.g. \cite{Mao}, \cite{Herb}, \cite{Dmitriev}, \cite{Xia}).

To take into account the relativistic kinematics of the correlations, many
authors simply replace $-k^{2}$ with $\omega ^{2}-k^{2}$ in the well known
non-relativistic expression derived by Migdal \cite{Mig71}. This is actually
incorrect and, as we show in our paper, results in the above problems in the
pion spectra. The completely relativistic accounting for the short-range
correlations in the pion self-energy was suggested recently in \cite{Lutz}.
In the following we calculate the relativistic short-range correlations in a
different way, and compare our result with that obtained in \cite{Lutz}. The
idea of this paper is to show how the relativistically correct treatment of
short-range correlation effects prevents the above-mentioned problems in the
pion spectra.

Since the short-range correlation distance is small as compared to the
Compton wavelength of the pion, we simulate the short-range correlations by
phenomenological contact terms in the Lagrangian density with Landau-Migdal
parameters $g_{NN}^{\prime }$, $g_{N\Delta }^{\prime }$, $g_{\Delta \Delta
}^{\prime }$, and derive the expression for the pion self-energy, which
retains the relativistic kinematics.

Since it is not our purpose to construct a fundamental renormalizable
theory, we consider the pseudovector coupling of pions with nucleons and
deltas, which is more likely to reproduce correctly the in-medium pion mass %
\cite{Serot}. We evaluate the pion self-energy by neglecting the vacuum
contribution, as widely used in the mean field approximation. In this case,
the only strongly model-dependent parameter arising in our calculations is
the effective nucleon mass, which, however, can be varied in order to
investigate the dependence of pion spectra on this parameter.

The paper is organized as follows. In Section 2 we briefly discuss the model
Lagrangian for the pion field in isosymmetric nucleon matter. In Section 3
we derive the relativistic form of the pion self-energy, including
short-range correlations in a nuclear medium, which we simulate by
relativistic phenomenological contact term in the Lagrangian. The expression
for the pion self-energy is obtained in terms of lowest-order polarization
tensors, which we calculate in Appendices A and B. In Sections 4, 5 we
discuss the pion dispersion relations. In Section 6 we present the result of
our numerical calculations for the pion spectra in isosymmetric nuclear
matter. A Summary and conclusions are in Section 7. In what follows, we use
the system of units $\hbar =c=1$. Summation over repeated Greek indices is
assumed.

\section{Formalism}

In isotopically symmetric nuclear matter, the dominant contribution to the
pion self-energy arises from particle-hole ($ph$) and $\Delta $-particle -
nucleon hole ($\Delta h$) excitations. The corresponding Lagrangian density
for the pion field can be written in the following form 
\begin{eqnarray}
\mathcal{L}_{\pi } &=&\frac{1}{2}\partial _{\mu }\mathbf{\pi }\partial ^{\mu
}\mathbf{\varphi }-\frac{1}{2}m_{\pi }^{2}\mathbf{\varphi }^{2}\mathbf{-}%
\frac{f}{m_{\pi }}\bar{\psi}_{N}\gamma ^{\mu }\gamma _{5}\mathbf{\tau }\psi
_{N}\partial _{\mu }\mathbf{\varphi }  \notag \\
&&+\frac{f_{\Delta }}{m_{\pi }}\bar{\psi}_{\Delta }^{\mu }\mathbf{T}^{+}\psi
_{N}\partial _{\mu }\mathbf{\varphi }+\frac{f_{\Delta }}{m_{\pi }}\bar{\psi}%
_{N}\mathbf{T}\psi _{\Delta }^{\mu }\partial _{\mu }\mathbf{\varphi }.
\label{lp}
\end{eqnarray}%
Here $\mathbf{\varphi }$ is the pseudoscalar isovector pion field, $m_{\pi }$
is the bare pion mass, and $f=0.988$ is the pion-nucleon coupling constant.
The excitation of the $\Delta $ in pion-nucleon scattering is described by
the last two terms in the Lagrangian with the Chew-Low value of the coupling
constant, $f_{\pi N\Delta }=2f$. The nucleon field is denoted as $\psi _{N}$%
, and $\psi _{\Delta }$ stands for the Rarita-Schwinger spinor of the $%
\Delta $-baryon. Here and below, we denote as $\mathbf{\tau }$ the isospin $%
1/2$ operators, which act on the isobaric doublet $\psi $ of the nucleon
field. The $\Delta $-barion is an isospin $3/2$ particle represented by a
quartet of four states. $\mathbf{T}$ is the isospin transition operator
between the isospin $1/2$ and $3/2$ fields.

The free-pion Green function is given by\footnote{%
The pion propagator is diagonal in isospin space, therefore we omit the
isospin indices.}%
\begin{equation}
D\left( K\right) =\frac{1}{\omega ^{2}-k^{2}-m_{\pi }^{2}+i0},  \label{free}
\end{equation}%
The in-medium pion propagator obeys the Dyson's equation%
\begin{equation}
\tilde{D}\left( K\right) =D\left( K\right) +D\left( K\right) \tilde{\Pi}%
\left( K\right) \tilde{D}\left( K\right) ,  \label{Dyson}
\end{equation}%
where the self-energy $\tilde{\Pi}\left( K\right) $ arises due to the pion
interactions with nucleons and delta-resonances. The four-momentum of a pion
is denoted as 
\begin{equation}
K=\left( \omega ,\mathbf{k}\right) ,  \label{Kmu}
\end{equation}%
and we use the notation $K_{\mu }^{2}=$ $\omega ^{2}-k^{2}$.

\section{Short-range correlations}

The $ph$ and $\Delta h$ contributions to the pion self-energy can be
evaluated via the diagramms in Fig. 1,

\vskip0.3cm

\psfig{file=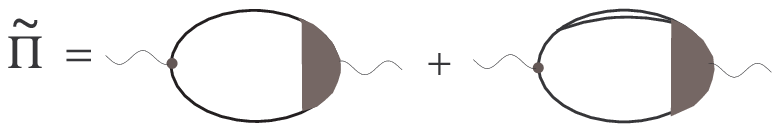} {Fig. 1. Diagrammatic representation for the pion
self-energy. The first graph corresponds to a virtual nucleon particle--hole
pair, and the second one is due to nucleon hole--$\Delta$-isobar excitations. The $\Delta$-isobar is
shown by a double-line. The shadowed effective vertices for the pion
interaction with nucleons and deltas account for short-range
correlations in the medium.}

\vskip0.3cm These graphs include the nucleon particle--hole pair and the
nucleon hole--$\Delta $-isobar pair. The shadowed effective vertices for the
pion interaction with nucleons and deltas take into account the short-range
correlations in the medium. These vertices are irreducible with respect to
pion lines. We perform the relativistic calculation by assuming that the
short-range correlation distance is small as compared to the Compton
wavelength of the pion. In this case, the correlations can be simulated by
phenomenological contact terms in the Lagrangian density of the form 
\begin{eqnarray}
\mathcal{L}_{\mathrm{corr}} &=&\frac{f^{2}}{m_{\pi }^{2}}g_{NN}^{\prime
}\left( \bar{\psi}\gamma _{\nu }\gamma _{5}\mathbf{\tau }\psi \right) \left( 
\bar{\psi}\gamma ^{\nu }\gamma _{5}\mathbf{\tau }\psi \right) +\frac{%
f_{\Delta }^{2}}{m_{\pi }^{2}}g_{\Delta \Delta }^{\prime }\left( \bar{\psi}%
_{\Delta }^{\mu }\mathbf{T}^{+}\psi _{N}\right) \left( \bar{\psi}_{N}\mathbf{%
T}\psi _{\Delta \mu }\right)  \notag \\
&&+\frac{ff_{\Delta }}{m_{\pi }^{2}}g_{N\Delta }^{\prime }\left( \bar{\psi}%
_{\Delta }^{\mu }\mathbf{T}^{+}\psi _{N}\right) \left( \bar{\psi}_{N}\gamma
_{\mu }\gamma _{5}\mathbf{\tau }\psi _{N}\right)  \label{SR}
\end{eqnarray}%
With the allowance of a derivative coupling of pions with nucleons and
deltas, the pion self-energy, shown in Fig. 1, can be written in the
following form%
\begin{equation}
\tilde{\Pi}=\tilde{\Pi}^{\mu \nu }K_{\mu }K_{\nu }  \label{Se}
\end{equation}%
where $\tilde{\Pi}^{\mu \nu }$ stands for the medium polarization tensor 
\begin{equation}
\tilde{\Pi}^{\mu \nu }=\tilde{\Pi}_{ph}^{\mu \nu }+\tilde{\Pi}_{\Delta
h}^{\mu \nu }  \label{PiTot}
\end{equation}%
which incorporates two terms, $\tilde{\Pi}_{ph}^{\mu \nu }$ and $\tilde{\Pi}%
_{\Delta h}^{\mu \nu }$, connected by the Dyson's equations depicted
graphically in Fig. 2. Here, the small rectangular blocks represent the
short-range interactions (\ref{SR}).

\vskip0.3cm

\psfig{file=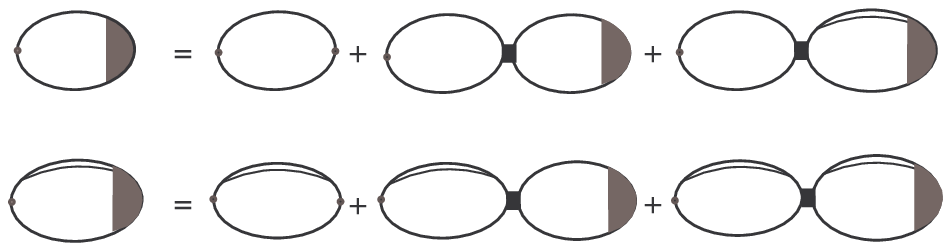} {Fig. 2. The diagrammatic set of the
Dyson's equations connecting the two terms shown in Fig. 1. The small
rectangular blocks represent the short-range interactions.}

\vskip0.3cm Corresponding to the diagrams of Fig. 2 we have the following
equations for the polarization tensors%
\begin{equation}
\tilde{\Pi}_{ph}^{\mu \nu }=\Pi _{ph}^{\mu \nu }-g_{NN}^{\prime }\Pi
_{ph}^{\mu \alpha }g_{\alpha \beta }\tilde{\Pi}_{ph}^{\beta \nu }-g_{N\Delta
}^{\prime }\Pi _{ph}^{\mu \alpha }g_{\alpha \beta }\tilde{\Pi}_{\Delta
h}^{\beta \nu }\,\,,  \label{DysN}
\end{equation}%
\begin{equation}
\tilde{\Pi}_{\Delta h}^{\mu \nu }=\Pi _{\Delta h}^{\mu \nu }-g_{N\Delta
}^{\prime }\Pi _{\Delta h}^{\mu \alpha }g_{\alpha \beta }\tilde{\Pi}%
_{ph}^{\beta \nu }-g_{\Delta \Delta }^{\prime }\Pi _{\Delta h}^{\mu \alpha
}g_{\alpha \beta }\tilde{\Pi}_{\Delta h}^{\beta \nu }\,\,,  \label{DysD}
\end{equation}%
where $g^{\rho \mu }=\mathsf{\mathrm{diag}}(1,-1,-1,-1)$ is the signature
tensor. The lowest-order polarizations $\Pi _{ph}^{\mu \nu }$ and $\Pi
_{\Delta h}^{\mu \nu }$ are defined as the following one-loop integrals 
\begin{equation}
\Pi _{ph}^{\mu \nu }\left( \omega ,\mathbf{k}\right) =-i\frac{2f^{2}}{m_{\pi
}^{2}}\limfunc{Tr}\int \frac{d^{4}p}{(2\pi )^{4}}\,\,\,{G}(p)\gamma ^{\mu
}\gamma _{5}{G}(p+K)\,\gamma ^{\nu }\gamma _{5},  \label{ph}
\end{equation}%
\begin{eqnarray}
\Pi _{\Delta h}^{\mu \nu } &=&-i\frac{f_{\pi N\Delta }^{2}}{m_{\pi }^{2}}%
\frac{4}{3}\int \frac{d^{4}p}{(2\pi )^{4}}\,\,\limfunc{Tr}{G}(p){S}^{\mu \nu
}(p_{\Delta }+K)\,\,  \notag \\
&&-i\frac{f_{\pi N\Delta }^{2}}{m_{\pi }^{2}}\frac{4}{3}\int \frac{d^{4}p}{%
(2\pi )^{4}}\,\,\limfunc{Tr}{G}(p){S}^{\mu \nu }(p_{\Delta }-K),  \label{dh}
\end{eqnarray}%
where ${G}(p)$ denotes the in-medium nucleon propagator, and ${S}^{\mu \nu
}\left( p\right) $ stands for the Rarita-Schwinger propagator for the
in-medium $\Delta $ particle.

The set of 32 equations (\ref{DysN}), (\ref{DysD}) for the components $%
\tilde{\Pi}_{ph}^{\mu \nu }$, $\tilde{\Pi}_{\Delta h}^{\mu \nu }$\ with $\mu
,\nu =0,1,2,3$ simplifies in the reference frame where $K=\left( \omega
,0,0,k\right) $. With taking also into account that the transverse (with
respect to $K$) components of the polarization tensor do not mix to the
longitudinal ones, the above equations reduce to 8 linear algebraic
equations for the components $\tilde{\Pi}_{ph}^{00}$, $\tilde{\Pi}_{ph}^{03}$%
, $\tilde{\Pi}_{ph}^{30}$, $\tilde{\Pi}_{ph}^{33}$, $\tilde{\Pi}_{\Delta
h}^{00}$, $\tilde{\Pi}_{\Delta h}^{03}$, $\tilde{\Pi}_{\Delta h}^{30}$, $%
\tilde{\Pi}_{\Delta h}^{33}$. Solution of the above equations and
contraction of the resulting tensor with $K_{\mu }K_{\nu }$, as given by Eq.
(\ref{Se}), gives the pion self energy in the following form 
\begin{eqnarray}
\tilde{\Pi} &=&\frac{1}{W}\left[ \left( 1+\left( g_{\Delta \Delta }^{\prime
}-g_{N\Delta }^{\prime }\right) \left( A_{\Delta }-\left( g_{\Delta \Delta
}^{\prime }-g_{N\Delta }^{\prime }\right) B_{\Delta }\right) \right) \Pi
_{ph}\right.  \notag \\
&&+\left( 1+\left( g_{NN}^{\prime }-g_{N\Delta }^{\prime }\right) \left(
A_{N}-\left( g_{NN}^{\prime }-g_{N\Delta }^{\prime }\right) B_{N}\right)
\right) \Pi _{\Delta h}  \notag \\
&&+K_{\mu }^{2}\left\{ \left( g_{N\Delta }^{\prime 2}-g_{NN}^{\prime
}g_{\Delta \Delta }^{\prime }\right) A_{N}B_{\Delta }-g_{\Delta \Delta
}^{\prime }B_{\Delta }\right.  \notag \\
&&+\left( g_{N\Delta }^{\prime 2}-g_{NN}^{\prime }g_{\Delta \Delta }^{\prime
}\right) A_{\Delta }B_{N}-g_{NN}^{\prime }B_{N}-g_{N\Delta }^{\prime
}\digamma _{N\Delta }  \notag \\
&&\left. \left. -\left( g_{NN}^{\prime }+g_{\Delta \Delta }^{\prime
}-2g_{N\Delta }^{\prime }\right) \left( g_{N\Delta }^{\prime
2}-g_{NN}^{\prime }g_{\Delta \Delta }^{\prime }\right) B_{N}B_{\Delta
}\right\} \right] ,  \label{dgen}
\end{eqnarray}%
where the following notations are used%
\begin{eqnarray}
W &=&1+g_{NN}^{\prime }A_{N}-g_{NN}^{\prime 2}B_{N}+g_{\Delta \Delta
}^{\prime }A_{\Delta }-g_{\Delta \Delta }^{\prime 2}B_{\Delta
}+g_{NN}^{\prime }g_{\Delta \Delta }^{\prime }A_{N}A_{\Delta }  \notag \\
&&+\left( g_{N\Delta }^{\prime 2}-g_{NN}^{\prime }g_{\Delta \Delta }^{\prime
}\right) \left( g_{\Delta \Delta }^{\prime }A_{N}\ B_{\Delta
}+g_{NN}^{\prime }A_{\Delta }B_{N}\right)  \notag \\
&&+\left( g_{N\Delta }^{\prime 2}-g_{NN}^{\prime }g_{\Delta \Delta }^{\prime
}\right) ^{2}B_{N}B_{\Delta }-g_{N\Delta }^{\prime 2}\Phi _{N\Delta },
\label{W}
\end{eqnarray}%
\begin{equation}
\Pi _{ph}=K_{\mu }\Pi _{ph}^{\mu \nu }K_{\nu },\,\ \ \ \ \ \ \Pi _{\Delta
h}=K_{\mu }\Pi _{\Delta h}^{\mu \nu }K_{\nu }  \label{n1}
\end{equation}%
\begin{equation}
A_{N}=\Pi _{ph}^{00}-\Pi _{ph}^{33},\,\ \ \ \ B_{N}=\Pi _{ph}^{00}\Pi
_{ph}^{33}-\Pi _{ph}^{03}\Pi _{ph}^{30},  \label{n2}
\end{equation}%
\begin{equation}
\ A_{\Delta }=\Pi _{\Delta h}^{00}-\Pi _{\Delta h}^{33},\,\ \ \ B_{\Delta
}=\Pi _{\Delta h}^{00}\Pi _{\Delta h}^{33}-\Pi _{\Delta h}^{03}\Pi _{\Delta
h}^{30}  \label{n3}
\end{equation}%
\begin{equation}
\digamma _{N\Delta }=\Pi _{ph}^{00}\Pi _{\Delta h}^{33}-\Pi _{ph}^{03}\Pi
_{\Delta h}^{30}-\Pi _{ph}^{30}\Pi _{\Delta h}^{03}+\Pi _{\Delta h}^{00}\Pi
_{ph}^{33}  \label{n4}
\end{equation}%
\begin{equation}
\Phi _{N\Delta }=\Pi _{ph}^{00}\Pi _{\Delta h}^{00}-\Pi _{ph}^{03}\Pi
_{\Delta h}^{30}-\Pi _{ph}^{30}\Pi _{\Delta h}^{03}+\Pi _{\Delta h}^{33}\Pi
_{ph}^{33}  \label{n5}
\end{equation}%
A standard calculation of the lowest order polarization tensors is simple
but results in cumbersome expressions, therefore we preferred to show them
in the Appendices A, B. The explicit relativistic expressions for the real
and imaginary part of the polarizations are given by Eqs. (\ref{r00}) - (\ref%
{r33}), (\ref{Imp1}) and (\ref{RD}), (\ref{ID}).

It is interesting to compare our Eq. (\ref{dgen}) with the corresponding
expression derived recently in \cite{Lutz}. As can be seen, the pion
self-energy given in \cite{Lutz} by Eq. (9), in the limit of vanishing
Landau-Migdal couplings, does not reproduce the lowest-order self-energy
given by Eq. (3) of the same work, having instead the opposite sign. Using
standard definitions for the polarization tensor, all the minus signs in Eq.
(9) of the work \cite{Lutz} should be replaced with plus. After these
modifications the result obtained in \cite{Lutz}, for the case of nuclear
matter at rest, becomes identical to our Eq. (\ref{dgen}).

It is instructive to show how the relativistic expression (\ref{dgen})
transforms into the known non-relativistic form in the limit $p_{n},p_{p}\ll
M^{\ast }$. The latter conditions assume also $k\ll M$ and $\omega \lesssim
p_{n}^{2}/\left( 2M^{\ast }\right) $ $\ll k$. In this case, the
non-relativistic reduction of the pion-nucleon and pion-delta coupling leads
to an effective interaction Hamiltonian of the form 
\begin{equation}
\mathcal{H}_{int}=\frac{f}{m_{\pi }}\left( \mathbf{\sigma }\cdot \mathbf{%
\nabla }\right) \left( \mathbf{\tau }\cdot \mathbf{\varphi }\right) +\frac{%
f_{\Delta }}{m_{\pi }}\left( \mathbf{S}^{+}\cdot \mathbf{\nabla }\right)
\left( \mathbf{T}^{+}\cdot \mathbf{\varphi }\right) +h.c.  \label{H}
\end{equation}%
where $\mathbf{S}^{+}$ and $\mathbf{T}^{+}$ are the transition spin and
isospin operators, respectively, connecting spin $1/2$ and $3/2$ states.
Thus, in the non-relativistic limit only the $\Pi _{ph}^{33}$ and $\Pi
_{\Delta h}^{33}$ components of the lowest-order polarization tensor
contribute to the pion self-energy, and the rest of components can be
neglected. This yields%
\begin{equation}
\tilde{\Pi}_{\mathrm{nr}}=\frac{k^{2}\left[ \left( \Pi _{ph}^{33}+\Pi
_{\Delta h}^{33}\right) -\left( g_{NN}^{\prime }+g_{\Delta \Delta }^{\prime
}-2g_{N\Delta }^{\prime }\right) \Pi _{ph}^{33}\Pi _{\Delta h}^{33}\right] }{%
1-g_{NN}^{\prime }\Pi _{ph}^{33}-g_{\Delta \Delta }^{\prime }\Pi _{\Delta
h}^{33}+\left( g_{NN}^{\prime }g_{\Delta \Delta }^{\prime }-g_{N\Delta
}^{\prime 2}\right) \Pi _{ph}^{33}\Pi _{\Delta h}^{33}}.  \label{Pinr}
\end{equation}
If, for the moment, one assumes the universal coupling for nucleons and
deltas, i.e. $g_{NN}^{\prime }=g_{N\Delta }^{\prime }=g_{\Delta \Delta
}^{\prime }=g^{\prime }$, then 
\begin{equation*}
\tilde{\Pi}_{\mathrm{nr}}\simeq \frac{k^{2}\Pi ^{33}}{1-g^{\prime }\Pi ^{33}}
\end{equation*}%
The $\Pi ^{33}$ component can be identified with the non-relativistic pion
susceptibility $\chi =-\Pi ^{33}$. Then we arrive to the well-known
non-relativistic form for the pion self-energy \cite{Eric}:%
\begin{equation}
\tilde{\Pi}_{\mathrm{nr}}\simeq \frac{-k^{2}\chi }{1+g^{\prime }\chi }
\label{eric}
\end{equation}%
where the non-relativistic pion susceptibility is given by (see Appendix A
and B) 
\begin{eqnarray}
\chi  &=&\frac{4f^{2}}{m_{\pi }^{2}}\left( \Phi _{0}\left( \omega
,k,p_{F}\right) +\Phi _{0}\left( -\omega ,k,p_{F}\right) \right)   \notag \\
&&-\frac{f_{\pi N\Delta }^{2}}{9m_{\pi }^{2}}\left( \Phi _{0}\left( \omega
-\omega _{R},k;p_{F}\right) -\Phi _{0}\left( -\omega -\omega
_{R},k;p_{F}\right) \right)   \label{hi}
\end{eqnarray}%
Here $\Phi _{0}\left( \omega ,k;p_{F}\right) $ is the Migdal function Eq. (%
\ref{MigF}), and the $\Delta $-resonant frequency is given by 
\begin{equation}
\omega _{R}=\frac{M_{\Delta }^{\ast 2}-M^{\ast 2}}{2M^{\ast }}.  \label{wr}
\end{equation}%
In the static limit, $M_{\Delta }^{\ast },M^{\ast }\rightarrow \infty $,
this expression can be reduced to the widely used form \cite{Eric}:%
\begin{equation}
\chi =\frac{f^{2}n}{m_{\pi }^{2}}\left( \frac{1}{k^{2}/2M^{\ast }-\omega }+%
\frac{1}{k^{2}/2M^{\ast }+\omega }\right) +\frac{8}{9}\frac{f_{\pi N\Delta
}^{2}}{m_{\pi }^{2}}\frac{n\omega _{R}}{\omega _{R}^{2}-\omega ^{2}},
\label{hist}
\end{equation}%
with $n=n_{n}+n_{p}$ being the total number density of nucleons.

\section{Pion-like excitations in nuclear matter}

The Dyson equation Eq. (\ref{Dyson}) for the pion propagator can be solved
to give%
\begin{equation}
\tilde{D}^{-1}\left( \omega ,k\right) =\omega ^{2}-m_{\pi }^{2}-k^{2}-\tilde{%
\Pi}\left( \omega ,k\right)  \label{prop}
\end{equation}%
where the pion self-energy is given by Eq. (\ref{dgen}). The eigen-modes of
the pion field in the medium are found from the poles of the pion
propagator. In general, the pion self-energy has both a real and an
imaginary part. From the poles of the propagator (\ref{prop}) we obtain the
following dispersion equation 
\begin{equation}
\left[ \omega ^{2}-m_{\pi }^{2}-k^{2}-\func{Re}\tilde{\Pi}\left( \omega
,k\right) \right] ^{2}+\left[ \func{Im}\tilde{\Pi}\left( \omega ,k\right) %
\right] ^{2}=0,  \label{disp}
\end{equation}%
which has no real solutions $\omega \left( k\right) $ if $\func{Im}\tilde{\Pi%
}\left( \omega ,k\right) \neq 0$. In this case the system has no stationary
excitations, although it possesses resonant properties at some frequencies.
The retarded pion propagator $\tilde{D}_{ab}^{R}\left( \omega ,k\right)
=\delta _{ab}\tilde{D}^{R}\left( \omega ,k\right) $ represents a generalized
susceptibility for the components $\varphi _{a}$ of the pion field.
According to the fluctuation-dissipation theorem, the spectral distribution
of the pion field fluctuations can be expressed as%
\begin{equation}
\left( \varphi _{a}^{2}\right) _{\omega ,\mathbf{k}}=\frac{i}{2}\coth \left( 
\frac{\omega }{2T}\right) \left[ \tilde{D}^{R}\left( \omega ,k\right) -%
\tilde{D}^{R\ast }\left( \omega ,k\right) \right]  \label{fl}
\end{equation}%
The imaginary part of the retarded pion propagator is connected to that of
the time-ordered (causal) propagator as%
\begin{equation}
\func{Im}\tilde{D}^{R}\left( \omega ,k\right) =\tanh \left( \frac{\omega }{2T%
}\right) \func{Im}\tilde{D}\left( \omega ,k\right) ,  \label{ImDR}
\end{equation}%
which results in%
\begin{equation}
\left( \varphi _{a}^{2}\right) _{\omega ,\mathbf{k}}=-\func{Im}\tilde{D}%
\left( \omega ,k\right)  \label{fluc}
\end{equation}%
Thus, the function 
\begin{equation}
-\func{Im}\tilde{D}\left( \omega ,k\right) =\frac{-\func{Im}\tilde{\Pi}%
\left( \omega ,k\right) }{\left[ \omega ^{2}-m_{\pi }^{2}-k^{2}-\func{Re}%
\tilde{\Pi}\left( \omega ,k\right) \right] ^{2}+\left[ \func{Im}\tilde{\Pi}%
\left( \omega ,k\right) \right] ^{2}}.  \label{gam}
\end{equation}%
contains a complete information about the pion field fluctuations. It
possesses maxima\footnote{%
For bosons, the imaginary part of the causal propagator is negative.} at the
frequencies corresponding to the pion-like eigen modes in the system.
Indeed, in the limiting case of $\func{Im}\tilde{\Pi}\left( \omega ,k\right)
\rightarrow 0$ one has $-\func{Im}\tilde{D}\left( \omega ,k\right)
\rightarrow \pi \delta \left( \omega ^{2}-m_{\pi }^{2}-k^{2}-\func{Re}\tilde{%
\Pi}\left( \omega ,k\right) \right) $, so that it differs from zero only if
the frequency $\omega \left( k\right) $ is a solution to the above
dispersion equation. If the imaginary part of the pion self-energy is a
small, but finite, value the function $-\func{Im}\tilde{D}\left( \omega
,k\right) $ is sharply peaked along a line $\omega \left( k\right) $ where 
\begin{equation}
\omega ^{2}-m_{\pi }^{2}-k^{2}-\func{Re}\tilde{\Pi}\left( \omega ,k\right) =0
\label{branch}
\end{equation}%
and the width $\gamma \left( k\right) $ of the resonance is connected to the
momentum-dependent decay rate of the excitation $\omega \left( k\right) $.
If the condition $\gamma \left( k\right) \ll \omega \left( k\right) $ is not
fulfilled, any small perturbation of frequency $\omega \left( k\right) $
undergoes aperiodic damping.

\section{Results}

To illuminate the problem of the relativistic pion spectra mentioned in the
Introduction we, at first, reproduce the widely used calculations, where the
''relativistic'' corrections to the short-range correlation are included by
the simple replacement of $-k^{2}$ with $\omega ^{2}-k^{2}$ in the
non-relativistic Eq. (\ref{eric}). As it has been made by many authors, we
perform this calculation by assuming the universality of the Landau-Migdal
coupling, $g_{NN}^{\prime }=g_{N\Delta }^{\prime }=g_{\Delta \Delta
}^{\prime }=g^{\prime }$.

\vskip0.3cm

\psfig{file=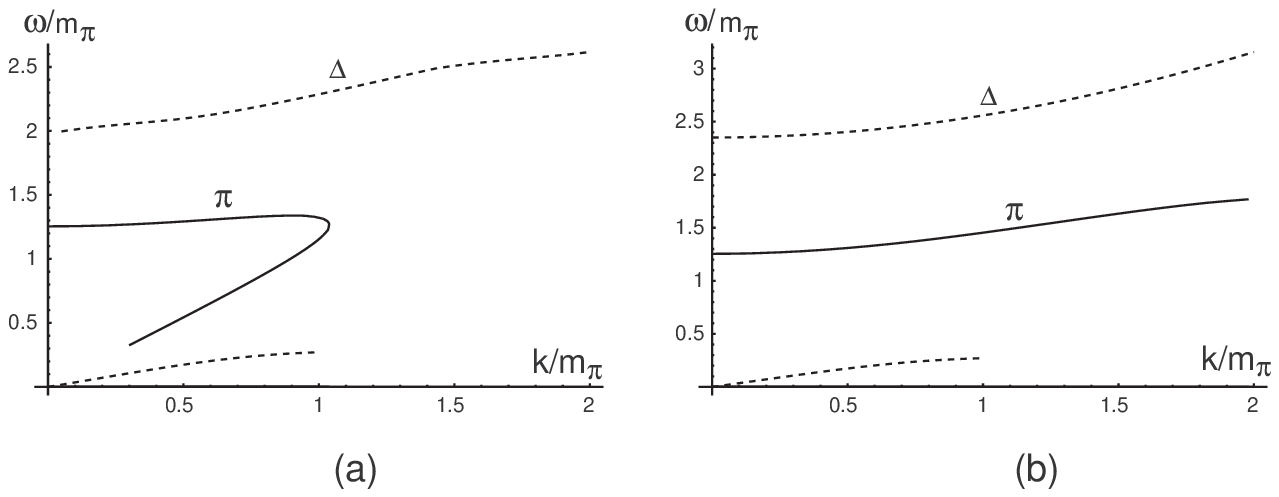} {Fig. 3. Pion spectra obtained with universal coupling 
$g^\prime =0.6$ for isosymmetric nuclear matter at saturation density $n=n_0$. 
(a) The spectrum obtained with incorrect inclusion of the short-range correlations. 
(b)  Same but now with inclusion of the correlation effects as given by 
Eq. (12)}

\vskip0.3cm

When the ''relativistic'' corrections to the short-range correlation are
included by the simple replacement of $-k^{2}$ with $\omega ^{2}-k^{2}$ in
the non-relativistic Eq. (\ref{eric}), we obtain the pion spectrum as shown
in Fig. 3a. This spectrum, obtained with $g^{\prime }=0.6$ at the saturation
density $n_{0}=0.17\,fm^{-3}$, possesses an unphysical behaviour, with two
pion-like branches merging at some point where the quasi-particle velocity, $%
d\omega /dk$, becomes infinite. In Fig. 3b we show the result of the same
calculation, but now incorporating the pion self-energy as given by Eq.(\ref%
{dgen}). This spectrum of pion-like excitations drastically differs from the
spectrum of Fig. 3a. Here the solid lines are solutions to Eq. (\ref{disp})
where the imaginary part of the self-energy vanishes. In the same picture,
the dashed lines show the solutions to Eq. (\ref{branch}) in the domains
with a non-zero imaginary part of the pion self-energy. The lower dashed
line represents the spectrum of the spin-isospin sound, which undergoes a
strong Landau damping. The above relativistic corrections to this
low-frequency mode are negligible, therefore it is of the same form as in
Fig. 3a. The upper dashed line is the $\Delta $-resonant mode.

As one can see, the obtained spectrum does not contain any pion branches
where the pion velocity becomes infinite, in contrast to the ones obtained
before in relativistic calculations of many authors. We claim that this
result is due to the correct relativistic incorporation of the short-range
correlations.

The experimental information about the Landau-Migdal parameters is very
limited. Previous theories basically used the universal ansatz, $%
g_{NN}^{\prime }=g_{N\Delta }^{\prime }=g_{\Delta \Delta }^{\prime
}=g^{\prime }$, with $g^{\prime }=0.6\div 0.7$. However, modern experiments
and theoretical estimates \cite{Wakasa}, \cite{Suzuki99}, \cite{Ari01} point
out that $g_{N\Delta }^{\prime }$ must be essentially smaller than $%
g_{NN}^{\prime }$ and $g_{\Delta \Delta }^{\prime }$. The most recent
analysis, reported in \cite{recent}, suggest $g_{NN}^{\prime }=0.6$, $%
g_{N\Delta }^{\prime }=0.24\pm 0.10$, $g_{\Delta \Delta }^{\prime }=0.6$.
While we will do not discuss possible deviations from this set of
Landau-Migdal parameters, let us investigate the behaviour of the pion
spectra in this case. First we calculate the pion spectrum for symmetric
nuclear matter at normal density by the use of the expression for the pion
self energy derived in \cite{Dmitriev}, \cite{Xia}, \cite{Herb} from the
non-relativistic form of the contact interaction, Eq. (\ref{H}). We use the
notation $\NEG{\Pi}$ for this incorrect form of the pion self-energy:%
\begin{equation}
\NEG{\Pi}=K_{\mu }^{2}\frac{\left[ \Pi _{ph}\left( K_{\mu }^{2}+g_{\Delta
\Delta }^{\prime }\Pi _{\Delta h}\right) +\Pi _{\Delta h}\left( K_{\mu
}^{2}+g_{NN}^{\prime }\Pi _{ph}\right) -2g_{N\Delta }^{\prime }\Pi _{ph}\Pi
_{\Delta h}\right] }{\left( K_{\mu }^{2}+g_{NN}^{\prime }\Pi _{ph}\right)
\left( K_{\mu }^{2}+g_{\Delta \Delta }^{\prime }\Pi _{\Delta h}\right)
-g_{N\Delta }^{\prime 2}\Pi _{ph}\Pi _{\Delta h}}  \label{Dmit}
\end{equation}%
The result of this calculation is shown in Fig. 4a, which explicitly
demonstrates the above-mentioned problem.

\vskip0.3cm

\psfig{file=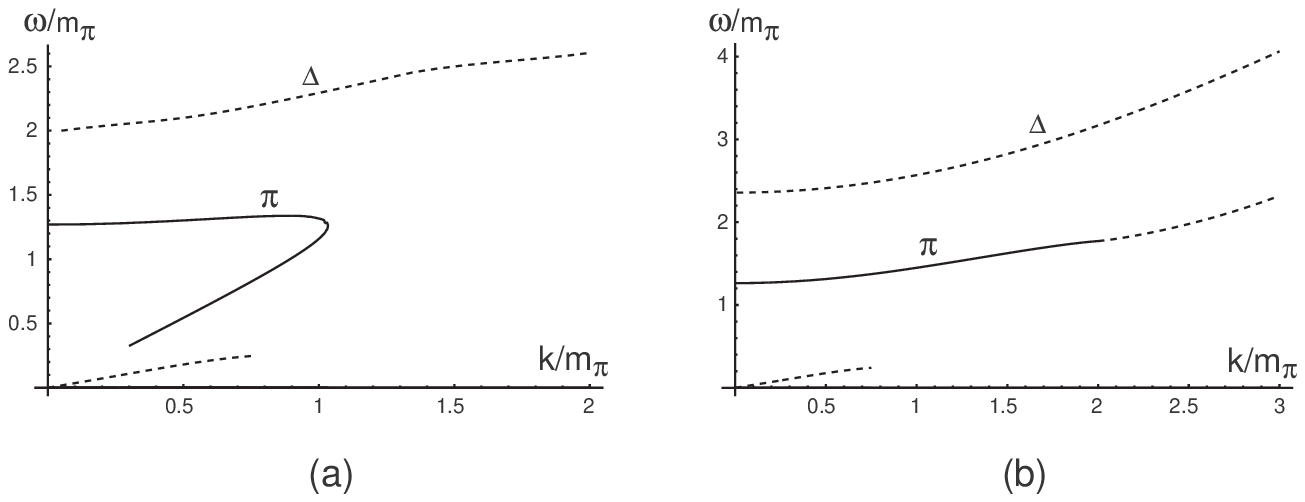} {Fig. 4. Pion spectra in symmetric nuclear matter at $n=n_0$ 
obtained with $g_{NN}^\prime = g_{\Delta\Delta}^\prime =0.6$ and 
$g_{N\Delta}^\prime =0.24$. (a) The short-range correlations are included as given by
Eq. (35). (b) The short-range correlations are included as given by
Eq. (12).}

\vskip0.3cm Two pion-like branches merge at some point where the
quasi-particle velocity, $d\omega /dk$, becomes infinite.

In Fig. 4b we show the pion spectrum obtained under the same conditions, but
with inclusion of the short-range correlations as given by our Eq. (\ref%
{dgen}). As in Fig. 3b, this spectrum does not contain any pion branches
where the pion velocity becomes infinite, in contrast to Fig. 4a. In order
to investigate the dependence of our results on the effective nucleon mass,
in Fig 5a, we show the curves obtained for the cases $M^{\ast }=0.7$ and $%
M^{\ast }=0.9$.

\vskip0.3cm

\psfig{file=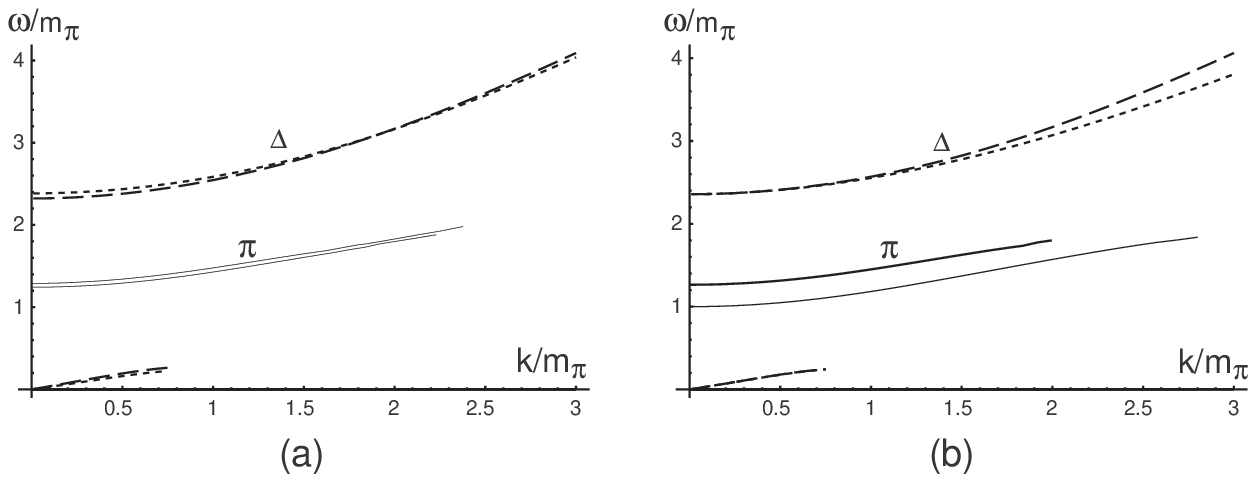} {Fig. 5. Pion spectra in symmetric nuclear matter at $n=n_0$ 
obtained with $g_{NN}^\prime = g_{\Delta\Delta}^\prime =0.6$ and 
$g_{N\Delta}^\prime =0.24$. (a) The curves obtained for the cases 
$M^{*}=0.7$ and $M^{*}=0.9$. In both cases the short-range correlations are 
included, as given by Eq. (12).
(b) The thick solid line and long-dash lines correspond to the relativistic calculation where the 
short-range correlations are included as in Eq. (12). The thin solid line and short-dashed 
lines are obtained with the non-relativistic accounting for the short-range correlations, 
as given by Eq. (20).}

\vskip0.3cm As one can see, the spectra obtained for these two cases are
very close to each other. The obtained curves are almost indistinguishable.

It is interesting to compare the pion spectrum shown in Fig. 4b with that
one obtained by the use of a simple non-relativistic expression, as given by
Eq. (\ref{Pinr}). In Fig. 5b we show the corresponding pion spectra. As can
be seen, the relativistic accounting for short-range correlations results in
an increasing in the effective pion mass, defined as $m_{\pi }^{\ast }=\sqrt{%
m_{\pi }^{2}+\tilde{\Pi}\left( m_{\pi }^{\ast },0\right) }$, by about 20\%.
This is caused by the delta-resonant contribution to the pion optical
potential, which becomes repulsive when $\omega >k$, due to relativistic
kinematics.

\section{Summary and Conclusion}

The purpose of this work was to demonstrate the crucial dependence of the
calculated pion spectra on the correct relativistic accounting for the
short-range correlation effects on the pion self-energy in the medium. We
have derived the pion self-energy in a nucleon medium with allowance for the
relativistic kinematics of short-range correlations. Since the short-range
correlation distance is small as compared to the Compton wavelength of the
pion, we simulated the short-range interactions by phenomenological contact
terms in the Lagrangian density with the Landau-Migdal parameters $%
g_{NN}^{\prime }$, $g_{N\Delta }^{\prime }$, $g_{\Delta \Delta }^{\prime }$.
The analytic expression for the pion self-energy is obtained in terms of the
lowest-order polarization tensors, which we have calculated on the basis of
the mean field ground state of the nucleon system.

In order to illuminate the effects of relativistic short-range correlations
we do not consider in our calculations any additional broadening like the
off-shell correction for the vertex form factors or the decay width for the $%
\Delta $-baryon in the medium.

We used the pion self-energy obtained in this way to solve the dispersion
equation for the pion-like excitations in cold symmetric nuclear matter at
the saturation density, and compare the obtained pion spectra with those
calculated by the use of the simplified expressions, which have been
employed by many authors.

As is clearly shown in Fig. 3a and in Fig. 4a, the pion spectra calculated
by the use of the simplified expressions possess an unphysical behaviour.
Similar spectra, which show the group velocity of a pion quasi-particle
larger than that of light, were obtained in relativistic calculations of
many authors, where the short-range correlations in nuclear matter are
incorporated in a simplified way. In Figs. 3b and 4b we show the pion
spectra calculated with the correct inclusion of relativistic effects, as
given by Eq. (\ref{dgen}). In this case the calculated pion branches
drastically differ from those of Figs. 3a and 4a and have a physically
meaningful behaviour. In order to investigate the dependence of the
calculated pion spectra on the model of nuclear matter, we repeated the
calculations by varying the effective nucleon mass. As shown in Fig. 5a, the
pion spectra, obtained with $M^{\ast }=0.7$ and $0.9$, are almost
indistinguishable.

We have also found that, due to relativistic kinematics, the delta-resonant
contribution to the pion optical potential is repulsive near the threshold.
This results in an increasing of the effective pion mass by about 20\%.

\section{Acknowledgements}

This work was carried out within the framework of the program of Presidium
RAS '' Non-stationary phenomena in astronomy'' and was partially supported
by Spanish grants FPA 2002-00612 and AYA 2001-3490-C02.

\appendix

\section{Lowest-order ph polarization}

In terms of the particle-hole propagators, the nucleon contribution to the
lowest-order polarization tensor consists on the direct and crossed terms,
as shown by loops in Fig. 6.

\vskip0.3cm

\psfig{file=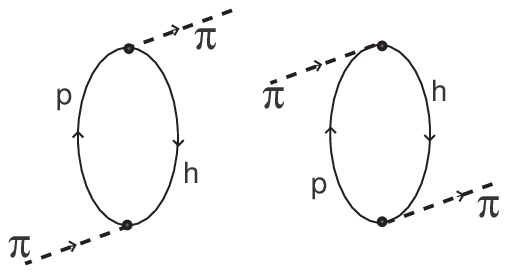} {Fig. 6. The direct and crossed contribution to
the $ph$ polarization tensor. }

\vskip0.3cm In the case of symmetric nuclear matter, the analytic expression
for polarization tensors are identical for the three pion species: 
\begin{equation}
\Pi _{ph}^{\mu \nu }\left( \omega ,\mathbf{k}\right) =-i\frac{2f^{2}}{m_{\pi
}^{2}}\limfunc{Tr}\int \frac{d^{4}p}{(2\pi )^{4}}\,\,\,{G}(p)\gamma ^{\mu
}\gamma _{5}{G}(p+q)\,\gamma ^{\nu }\gamma _{5}.  \label{PiNNt}
\end{equation}%
Here the nucleon propagator at finite density is given by 
\begin{equation}
{G}(p)=\left( \NEG{P}+M^{\ast }\right) \left[ \frac{1}{P^{2}-M^{\ast
2}+i\eta }+\frac{i\pi }{\varepsilon \left( \mathbf{p}\right) }\delta \left(
P^{0}-\varepsilon \left( \mathbf{p}\right) \right) f\left( \mathbf{p}\right) %
\right]  \label{GN}
\end{equation}%
with%
\begin{equation}
P=\left( p^{0}-U_{N},\mathbf{p}\right) ,  \label{Pnp}
\end{equation}%
$U_{N}$ is the nucleon mean-field potential. For isosymmetric nuclear
matter, one has $U_{p}=U_{n}$. Hereinafter we use the notation%
\begin{equation}
\varepsilon \left( \mathbf{p}\right) =\sqrt{M^{\ast 2}+\mathbf{p}^{2}}.
\label{eps}
\end{equation}%
The last term in Eq. (\ref{GN}), where $f\left( \mathbf{p}\right) $ is the
Fermi distribution for nucleons, $N=\left( n,p\right) $, corrects the free
nucleon propagator (Feynman piece) to account for the presence of the Femi
sea. To obtain the $ph$ contribution to the polarization we omit the product
of two Feynmann pieces when inserting the nucleon propagators into Eq. (\ref%
{PiNNt}). This is consistent with the mean-field approximation for the
ground state. After shifting the integration variable $p_{0}\rightarrow
p_{0}+U_{N}$ the density dependent part of the polarization tensor takes the
following form:%
\begin{eqnarray}
\Pi _{ph}^{\mu \nu }\left( K_{0},\mathbf{k}\right) ^{\,} &=&\frac{2f^{2}}{%
m_{\pi }^{2}}\int \frac{d^{4}p}{(2\pi )^{4}}\,\limfunc{Tr}\left( \NEG%
{p}+M^{\ast }\right) \gamma ^{\mu }\gamma _{5}\left( \NEG{p}+\NEG{K}+M^{\ast
}\right) \gamma ^{\nu }\gamma _{5}  \notag \\
&&\times \left[ i\pi ^{2}\frac{\delta \left( p^{0}-\varepsilon \left( 
\mathbf{p}\right) \right) f\left( \mathbf{p}\right) }{\varepsilon \left( 
\mathbf{p}\right) }\frac{\delta \left( p^{0}+\omega -\varepsilon \left( 
\mathbf{p+k}\right) \right) f\left( \mathbf{p+k}\right) }{\varepsilon \left( 
\mathbf{p+k}\right) }\right.  \notag \\
&&+\frac{\pi \delta \left( p^{0}-\varepsilon \left( \mathbf{p}\right)
\right) f\left( \mathbf{p}\right) }{\varepsilon \left( \mathbf{p}\right)
\left( \left( p+K\right) ^{2}-M^{\ast 2}+i\eta \right) }  \notag \\
&&\left. +\frac{\pi \delta \left( p^{0}+\omega -\varepsilon \left( \mathbf{%
p+k}\right) \right) f\left( \mathbf{p+k}\right) }{\varepsilon \left( \mathbf{%
p+k}\right) \left( p^{2}-M^{\ast 2}+i\eta \right) }\right] .  \label{PipNN}
\end{eqnarray}%
As it is apparent from the above equation, the polarization does not depend
on the value of the mean field potential.

In the following we need only the $\Pi _{ph}^{00}$, $\Pi _{ph}^{03}$, $\Pi
_{ph}^{30}$, $\Pi _{ph}^{33}$ components of the polarization tensor. The
zero-temperature Fermi distribution for nucleons, $f\left( \mathbf{p}\right)
=\theta \left( p_{F}-\left| \mathbf{p}\right| \right) $, is given by the
Heaviside's step-function, where $p_{n}=p_{p}=p_{F}$ stands for the nucleon
Fermi momentum. A standard calculation gives:%
\begin{eqnarray}
\func{Re}\Pi _{ph}^{00} &=&-\frac{f^{2}}{4\pi ^{2}}\frac{1}{m_{\pi }^{2}k}%
\int_{0}^{p_{F}}\frac{dpp}{\varepsilon }\left[ \left( 4p^{2}+K_{\mu
}^{2}\right) \ln \left| \frac{\left( K_{\mu }^{2}-2kp\right) ^{2}-4\omega
^{2}\varepsilon ^{2}}{\left( K_{\mu }^{2}+2kp\right) ^{2}-4\omega
^{2}\varepsilon ^{2}}\right| \right.  \notag \\
&&\left. +8kp+4\omega \varepsilon \ln \left| \frac{\left( K_{\mu
}^{2}\right) ^{2}-\left( 2kp-2\omega \varepsilon \right) ^{2}}{\left( K_{\mu
}^{2}\right) ^{2}-\left( 2kp+2\omega \varepsilon \right) ^{2}}\right| \right]
\label{r00}
\end{eqnarray}%
\begin{eqnarray}
\func{Re}\Pi _{ph}^{03} &=&\func{Re}\Pi _{ph}^{30}  \notag \\
&=&-\frac{f^{2}}{4\pi ^{2}}\frac{\omega }{m_{\pi }^{2}k^{2}}\int_{0}^{p_{F}}%
\frac{dpp}{\varepsilon }\left[ \left( 4\varepsilon ^{2}+K_{\mu }^{2}\right)
\ln \left| \frac{\left( K_{\mu }^{2}-2kp\right) ^{2}-4\omega ^{2}\varepsilon
^{2}}{\left( K_{\mu }^{2}+2kp\right) ^{2}-4\omega ^{2}\varepsilon ^{2}}%
\right| \right.  \notag \\
&&\left. +8kp+4\omega \varepsilon \ln \left| \frac{\left( K_{\mu
}^{2}\right) ^{2}-\left( 2kp-2\omega \varepsilon \right) ^{2}}{\left( K_{\mu
}^{2}\right) ^{2}-\left( 2kp+2\omega \varepsilon \right) ^{2}}\right| \right]
,  \label{r03}
\end{eqnarray}%
\begin{eqnarray}
\func{Re}\Pi _{ph}^{33} &=&-\frac{f^{2}}{4\pi ^{2}}\frac{\omega ^{2}}{m_{\pi
}^{2}k^{3}}\int_{0}^{p_{F}}\frac{dpp}{\varepsilon }  \notag \\
&&\times \left[ \left( \left( 4\varepsilon ^{2}+K_{\mu }^{2}\right) +4\frac{%
k^{2}}{\omega ^{2}}M^{2}\right) \ln \left| \frac{\left( K_{\mu
}^{2}-2kp\right) ^{2}-4\omega ^{2}\varepsilon ^{2}}{\left( K_{\mu
}^{2}+2kp\right) ^{2}-4\omega ^{2}\varepsilon ^{2}}\right| \right.  \notag \\
&&\left. +8pk+4\omega \varepsilon \ln \left| \frac{\left( K_{\mu
}^{2}\right) ^{2}-\left( 2kp-2\omega \varepsilon \right) ^{2}}{\left( K_{\mu
}^{2}\right) ^{2}-\left( 2kp+2\omega \varepsilon \right) ^{2}}\right| \right]
.  \label{r33}
\end{eqnarray}

The imaginary part of (\ref{PipNN}) is evaluated as%
\begin{eqnarray}
\func{Im}\Pi _{ph}^{\mu \nu } &=&-\frac{f^{2}}{m_{\pi }^{2}}\frac{1}{16\pi
^{2}}\int \,d^{4}p\limfunc{Tr}\left[ \left( \NEG{p}+M_{n}^{\ast }\right)
\gamma ^{\mu }\gamma _{5}\left( \NEG{p}+\NEG{K}+M^{\ast }\right) \gamma
^{\nu }\gamma _{5}\right]  \notag \\
&&\times \frac{\Theta \left( \varepsilon \left( \mathbf{p}\right) ,\omega
\right) }{\varepsilon \left( \mathbf{p}\right) \varepsilon \left( \mathbf{p+k%
}\right) }\delta \left( p^{0}-\varepsilon \left( \mathbf{p}\right) \right)
\delta \left( p^{0}+\omega -\varepsilon \left( \mathbf{p+k}\right) \right)
\label{Imp}
\end{eqnarray}%
where%
\begin{equation}
\Theta \left( \varepsilon ,\omega \right) \equiv f\left( \varepsilon \right) 
\left[ 1-f\left( \varepsilon +\omega \right) \right] +f\left( \varepsilon
+\omega \right) \left[ 1-f\left( \varepsilon \right) \right]  \label{theta}
\end{equation}%
At zero temperature, for the case $K_{\mu }^{2}<4M^{\ast 2}$ of our
interest, the imaginary part of the tensor is found to be%
\begin{eqnarray}
\func{Im}\Pi _{ph}^{\mu \nu } &=&-\frac{f^{2}}{12\pi }\frac{1}{m_{\pi }^{2}}%
\frac{1}{k}\theta \left( -K_{\mu }^{2}\right)  \notag \\
&&\times \left[ \theta \left( \omega \right) \theta \left( \varepsilon
_{F}-\varepsilon _{0}\right) J^{\mu \nu }\left( \max \left( \varepsilon
_{F}-\omega ,\varepsilon _{0}\right) ,\varepsilon _{F}\right) \right.  \notag
\\
&&\left. +\theta \left( -\omega \right) \theta \left( \varepsilon
_{F}-\omega -\varepsilon _{0}\right) J^{\mu \nu }\left( \max \left(
\varepsilon _{F},\varepsilon _{0}\right) ,\varepsilon _{F}-\omega \right) 
\right]  \label{Imp1}
\end{eqnarray}%
$\allowbreak $ $\allowbreak $ with%
\begin{eqnarray}
J^{00}\left( \varepsilon _{1},\varepsilon _{2}\right) &=&\left( \varepsilon
_{2}-\varepsilon _{1}\right) \left[ 4\left( \varepsilon _{1}^{2}+\varepsilon
_{2}\varepsilon _{1}+\varepsilon _{2}^{2}\right) \right.  \notag \\
&&\left. +6\omega \left( \varepsilon _{1}+\varepsilon _{2}\right) +3K_{\mu
}^{2}-12M^{\ast 2}\right] ,  \label{Imp2}
\end{eqnarray}%
\begin{eqnarray}
J^{03}\left( \varepsilon _{1},\varepsilon _{2}\right) &=&J^{30}\left(
\varepsilon _{1},\varepsilon _{2}\right) =\frac{\omega }{k}\left(
\varepsilon _{2}-\varepsilon _{1}\right) \left[ 4\left( \varepsilon
_{1}^{2}+\varepsilon _{2}\varepsilon _{1}+\varepsilon _{2}^{2}\right) \right.
\notag \\
&&\left. +6\omega \left( \varepsilon _{1}+\varepsilon _{2}\right) +3K_{\mu
}^{2}\right] ,  \label{Imp3}
\end{eqnarray}%
\begin{eqnarray}
J^{33}\left( \varepsilon _{1},\varepsilon _{2}\right) &=&\frac{\omega ^{2}}{%
k^{2}}\left( \varepsilon _{2}-\varepsilon _{1}\right) \left[ 4\left(
\varepsilon _{1}^{2}+\varepsilon _{2}\varepsilon _{1}+\varepsilon
_{2}^{2}\right) \right.  \notag \\
&&\left. +6\omega \left( \varepsilon _{1}+\varepsilon _{2}\right) +3K_{\mu
}^{2}+12M^{\ast 2}\frac{k^{2}}{\omega ^{2}}\right] .  \label{Imp4}
\end{eqnarray}%
In the above $\varepsilon _{0}=\sqrt{p_{\min }^{2}+M^{\ast 2}}$, where%
\begin{equation}
p_{\min }=\left| \frac{\omega }{2}\sqrt{1-\frac{4M^{\ast 2}}{K_{\mu }^{2}}}-%
\frac{k}{2}\right| ,  \label{pmin}
\end{equation}%
is the minimal space-momentum of a nucleon hole arising from the kinematical
restrictions of the process.

As one can easily check, the imaginary part of the polarization functions
equals to zero at $\omega =0$, and near this point behave as%
\begin{eqnarray}
\func{Im}\Pi _{ph}^{00} &\simeq &-\frac{f^{2}}{4\pi }\frac{1}{m_{\pi }^{2}}%
\left( 4p_{F}^{2}-k^{2}\right) \frac{\omega }{k}\func{sign}\left( \omega
\right)  \notag \\
\func{Im}\Pi _{ph}^{03} &\simeq &\func{Im}\Pi _{N\bar{N}}^{30}=-\frac{f^{2}}{%
4\pi }\frac{1}{m_{\pi }^{2}}\frac{\omega ^{2}}{k^{2}}\left( 4\varepsilon
_{F}^{2}-k^{2}\right) \func{sign}\left( \omega \right)  \notag \\
\func{Im}\Pi _{ph}^{33} &\simeq &-\frac{f^{2}}{4\pi }\frac{4M^{\ast 2}}{%
m_{\pi }^{2}}\frac{\omega }{k}\func{sign}\left( \omega \right)  \label{Sim}
\end{eqnarray}

By contracting the tensor $\Pi _{ph}^{\mu \nu }$ with $K_{\mu }K_{\nu }$ we
obtain the lowest-order pion self-energy caused by the nucleon-particle --
nucleon-hole excitations in the medium 
\begin{equation}
\func{Re}\Pi _{ph}\left( K\right) =\frac{f^{2}}{\pi ^{2}}\frac{K_{\mu
}^{2}M^{\ast 2}}{m_{\pi }^{2}k}\int_{0}^{p_{F}}\frac{dpp}{\varepsilon \left(
p\right) }\ln \left| \frac{\left( K_{\mu }^{2}-2kp\right) ^{2}-4\omega
^{2}\varepsilon ^{2}}{\left( K_{\mu }^{2}+2kp\right) ^{2}-4\omega
^{2}\varepsilon ^{2}}\right|  \label{SeNN}
\end{equation}%
This expression coincides with that obtained in \cite{Herb}. For the
imaginary part we have 
\begin{eqnarray}
\func{Im}\Pi _{ph} &=&\frac{f^{2}}{\pi }\frac{M^{\ast 2}}{m_{\pi }^{2}}\frac{%
K_{\mu }^{2}}{k}\theta \left( -K_{\mu }^{2}\right)  \notag \\
&&\times \left[ \theta \left( \omega \right) \theta \left( \varepsilon
_{F}-\varepsilon _{0}\right) \left( \varepsilon _{F}-\max \left( \varepsilon
_{F}-\omega ,\varepsilon _{0}\right) \right) \right.  \notag \\
&&\left. +\theta \left( -\omega \right) \theta \left( \varepsilon
_{F}-\omega -\varepsilon _{0}\right) \left( \varepsilon _{F}-\omega -\max
\left( \varepsilon _{F},\varepsilon _{0}\right) \right) \right]  \label{iph}
\end{eqnarray}%
As it follows from Eq. (\ref{Sim}), in the vicinity of $\omega =0$ one has%
\begin{equation}
\func{Im}\Pi _{ph}\simeq -\frac{f^{2}}{\pi }\frac{M^{\ast 2}}{m_{\pi }^{2}}%
k\omega \func{sign}\left( \omega \right)  \label{SimPi}
\end{equation}

It is instructive to consider the low-density limit of the function (\ref%
{SeNN}) in order to compare it with the well-known non-relativistic
expression. At low density of the nucleons, $p_{n,p}/M^{\ast }\ll 1$ one has 
$\varepsilon \left( p\right) \simeq M^{\ast }$. With this replacement, the
integration can be performed to give 
\begin{equation}
\func{Re}\Pi _{ph}\left( \omega ,k\right) =\frac{4f^{2}}{m_{\pi }^{2}}%
\,K_{\mu }^{2}\left( \Phi _{0}\left( \omega ,k;p_{F}\right) +\Phi _{0}\left(
-\omega ,k;p_{F}\right) \right) ,  \label{rph}
\end{equation}%
where 
\begin{equation}
\Phi _{0}\left( \omega ,k;p_{F}\right) =\frac{1}{4\pi ^{2}}\frac{M^{\ast 3}}{%
k^{3}}\left( \frac{1}{2}\left( a^{2}-k^{2}V_{F}^{2}\right) \ln \frac{a+kV_{F}%
}{a-kV_{F}}-akV_{F}\right)  \label{MigF}
\end{equation}%
is the Migdal function, with 
\begin{equation}
a=\omega +\frac{K_{\mu }^{2}}{2M^{\ast }},\ \ \ \ \ V_{F}=p_{F}/M^{\ast }
\label{av}
\end{equation}%
This non-relativistic limit of the particle-hole self-energy has been
obtained from relativistic kinematics. If one neglects the relativistic
kinematics, i.e. $K_{\mu }^{2}\rightarrow -k^{2}$, it reduces to the
standard non-relativistic formula stemming from the particle-hole excitation %
\cite{Migdal}, \cite{Eric}.

\section{Lowest order $\Delta h$ polarization}

The lowest order pion self energy involves also the terms caused by the
intrinsic excitation of a nucleon from below the Fermi surface into a $%
\Delta \left( 1232\right) $, as shown in Fig. 7.

\vskip0.3cm

\psfig{file=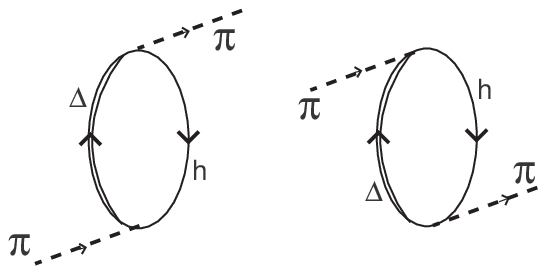} {Fig. 7. The direct and crossed contribution to
the $\Delta h$ polarization tensor.}

\vskip0.3cm Analytically we obtain%
\begin{eqnarray}
\Pi _{\Delta h}^{\mu \nu } &=&-i\frac{f_{\pi N\Delta }^{2}}{m_{\pi }^{2}}%
\frac{4}{3}\int \frac{d^{4}p}{(2\pi )^{4}}\,\,\limfunc{Tr}{G}_{D}(P_{n}){S}%
^{\mu \nu }(P_{\Delta }+K)\,\,  \notag \\
&&-i\frac{f_{\pi N\Delta }^{2}}{m_{\pi }^{2}}\frac{4}{3}\int \frac{d^{4}p}{%
(2\pi )^{4}}\,\,\limfunc{Tr}{G}_{D}(P_{n}){S}^{\mu \nu }(P_{\Delta }-K)
\label{Dn}
\end{eqnarray}%
where%
\begin{eqnarray}
S^{\mu \nu } &=&\frac{\left( \NEG{P}_{\Delta }+M_{\Delta }^{\ast }\right) }{%
P_{\Delta }^{2}-M_{\Delta }^{\ast 2}+i0}\left[ g_{\mu \nu }-\frac{1}{3}%
\gamma ^{\mu }\gamma ^{\nu }-\frac{2}{3M_{\Delta }^{\ast 2}}P_{\Delta }^{\mu
}P_{\Delta }^{\nu }-\frac{1}{3M_{\Delta }^{\ast }}\left( \gamma ^{\mu
}P_{\Delta }^{\nu }-\gamma ^{\nu }P_{\Delta }^{\mu }\right) \right] ,  \notag
\\
&&  \label{Smn}
\end{eqnarray}%
with $P_{\Delta }=\left( p_{0}-U_{\Delta },\mathbf{p}\right) $, denotes the
Rarita-Schwinger propagator for a $\Delta $ particle of effective mass $%
M_{\Delta }^{\ast }$. As widely used we employ the universal coupling of
mesons with nucleons and deltas, which is compatible with the data on
electromagnetic excitations of the $\Delta $ in nuclei \cite{Werh89}. In
this case the mean-field potentials for nucleons and deltas coincide, $%
U_{\Delta }=U_{N}$, and the effective mass of the $\Delta $ particle is
shifted by the scalar $\sigma $-field in the same way as that for nucleons
(see e.g. \cite{Herb}). We have%
\begin{equation}
M_{\Delta }^{\ast }=M_{\Delta }-M+M^{\ast }.  \label{MDel}
\end{equation}

This allows us to make a shift on the variable $p_{0}\rightarrow p_{0}+U_{N}$
to recast Eq. (\ref{Dn}) in the following form%
\begin{eqnarray}
\Pi _{\Delta h}^{\mu \nu } &=&\frac{f_{\pi N\Delta }^{2}}{m_{\pi }^{2}}\frac{%
4}{3}\int \frac{d^{4}p}{16\pi ^{3}}\,\,\limfunc{Tr}\left[ \left( \NEG%
{p}+M^{\ast }\right) {S}^{\mu \nu }(p+K)\right] \,\frac{\delta \left(
p_{0}-\varepsilon \left( \mathbf{p}\right) \right) f\left( \mathbf{p}\right) 
}{\varepsilon \left( \mathbf{p}\right) }\,  \notag \\
&&+\frac{f_{\pi N\Delta }^{2}}{m_{\pi }^{2}}\frac{4}{3}\int \frac{d^{4}p}{%
16\pi ^{3}}\,\,\limfunc{Tr}\left[ \left( \NEG{p}+M^{\ast }\right) {S}^{\mu
\nu }(p-K)\right] \frac{\delta \left( p_{0}-\varepsilon \left( \mathbf{p}%
\right) \right) f\left( \mathbf{p}\right) }{\varepsilon \left( \mathbf{p}%
\right) }.  \notag \\
&&  \label{PiDn}
\end{eqnarray}%
A calculation of the traces gives%
\begin{eqnarray}
&&\limfunc{Tr}\left( \NEG{p}+M^{\ast }\right) {S}^{\mu \nu }(p\pm K)  \notag
\\
&=&\frac{8}{3M_{\Delta }^{\ast 2}}\,\,\frac{\left( M^{\ast }M_{\Delta
}^{\ast }+M^{\ast 2}\pm Kp\right) }{\left( p\pm K\right) ^{2}-M_{\Delta
}^{\ast 2}+i0}\left( M_{\Delta }^{\ast 2}g^{\mu \nu }-(p^{\mu }\pm K^{\mu
})(p^{\nu }\pm K^{\nu })\right) .  \notag \\
&&  \label{trace}
\end{eqnarray}%
Here one can make the following remark. The problem of describing a spin $%
3/2 $ particle in relativistic field theory remains unsolved. Strictly
speaking, the expression (\ref{Smn}) is valid only for ''on-shell''
particles. Fortunately, due to the large effective mass of the $\Delta $%
-baryon, we do not need the totally relativistic form of Eq. (\ref{trace}).
This can be substantially simplified if we are interested in values of $%
\left| \omega \right| ,k\lesssim m_{\pi }$. By neglecting the small terms $%
\mathbf{kp\lesssim p}^{2}\sim p_{F}^{2}\ll M_{\Delta }^{\ast 2}$ in the
numerator of Eq. (\ref{trace}) we have the simpler form, which however
retains the relativistic kinematics. In this way we obtain%
\begin{equation}
\func{Re}\Pi _{\Delta h}^{33}\left( \omega ,k\right) =\frac{4f_{\pi N\Delta
}^{2}}{18}\frac{\left( M_{\Delta }^{\ast }+M^{\ast }\right) }{m_{\pi
}^{2}M^{\ast }}F_{\Delta }^{\left( 1\right) }\left( \omega ,k;p_{F}\right)
\,,\,\,\,\,  \label{RD}
\end{equation}%
\begin{equation}
\func{Re}\Pi _{\Delta h}^{03}\left( \omega ,k\right) =\func{Re}\Pi _{\Delta
h}^{30}\left( \omega ,k\right) =\frac{4f_{\pi N\Delta }^{2}}{18}\frac{%
kM^{\ast }}{M_{\Delta }^{\ast 2}}\frac{M_{\Delta }^{\ast }+M^{\ast }}{m_{\pi
}^{2}M^{\ast }}F_{\Delta }^{\left( 1\right) }\left( \omega ,k;p_{F}\right)
\,,\,\,\,\,  \label{RD03}
\end{equation}%
\begin{eqnarray}
\func{Re}\Pi _{\Delta h}^{00}\left( \omega ,k\right) &=&\frac{4f_{\pi
N\Delta }^{2}}{18}\frac{\left( M^{\ast }-M_{\Delta }^{\ast }\right) \left(
M_{\Delta }^{\ast }+M^{\ast }\right) ^{2}}{m_{\pi }^{2}M^{\ast }M_{\Delta
}^{\ast 2}}F_{\Delta }^{\left( 1\right) }\left( \omega ,k;p_{F}\right) 
\notag \\
&&+\frac{4f_{\pi N\Delta }^{2}}{18}\frac{\omega \left( M^{\ast }+M_{\Delta
}^{\ast }\right) ^{2}}{m_{\pi }^{2}M^{\ast }M_{\Delta }^{\ast 2}}F_{\Delta
}^{\left( 2\right) }\left( \omega ,k;p_{F}\right) \,,\,\,\,\,  \label{RD00}
\end{eqnarray}%
where the functions $F_{\Delta }^{\left( 1,2\right) }$ are defined as%
\begin{equation}
F_{\Delta }^{\left( 1\right) }\left( \omega ,k;p_{F}\right) =\frac{M^{\ast 2}%
}{\pi ^{2}k}\int_{0}^{p_{F}}dp\frac{p}{\varepsilon }\ln \left| \frac{\left(
M^{\ast }\omega _{R}-kp\right) ^{2}-\omega ^{2}\varepsilon ^{2}}{\left(
M^{\ast }\omega _{R}+kp\right) ^{2}-\omega ^{2}\varepsilon ^{2}}\right| ,
\label{FD}
\end{equation}%
\begin{equation*}
F_{\Delta }^{\left( 2\right) }\left( \omega ,k;p_{F}\right) =\frac{M^{\ast 2}%
}{\pi ^{2}k}\int_{0}^{p_{F}}dp\frac{p}{\varepsilon }\ln \left| \frac{\left(
\omega \varepsilon -kp\right) ^{2}-M^{\ast }\omega _{R}}{\left( \omega
\varepsilon +kp\right) ^{2}-M^{\ast }\omega _{R}}\right|
\end{equation*}%
and the $\Delta $-resonant frequency is given by 
\begin{equation}
\omega _{R}=\frac{M_{\Delta }^{\ast 2}-M^{\ast 2}-K_{\mu }^{2}}{2M^{\ast }}.
\label{wrr}
\end{equation}%
The imaginary part of the polarization arises from the poles of the
integrand of (\ref{PiDn}). It is non-vanishing for $0<K_{\mu }^{2}\leq
\omega _{R}^{2}$. For example, $\func{Im}\Pi _{\Delta h}^{33}$ is given by%
\begin{eqnarray}
\func{Im}\Pi _{\Delta h}^{33}\left( \omega ,k\right) &=&-\frac{2f_{\pi
N\Delta }^{2}}{9\pi }\frac{M^{\ast }M_{\Delta }^{\ast }+M^{\ast 2}}{m_{\pi
}^{2}k}\theta \left( \varepsilon _{2}\left( \omega ,k\right) -\varepsilon
_{1}\left( \omega ,k\right) \right)  \notag \\
&&\times \left( \varepsilon _{2}\left( \omega ,k\right) -\varepsilon
_{1}\left( \omega ,k\right) +\varepsilon _{2}\left( -\omega ,k\right)
-\varepsilon _{1}\left( -\omega ,k\right) \right) \func{sign}\left( \omega
\right)  \notag \\
&&  \label{ID}
\end{eqnarray}%
where%
\begin{equation}
\varepsilon _{1}\left( \omega ,k\right) =\max \left\{ M^{\ast },\frac{%
M^{\ast }}{K_{\mu }^{2}}\left( \omega \,\omega _{R}-\sqrt{k^{2}\left( \omega
_{R}^{2}-K_{\mu }^{2}\right) }\right) \right\}  \label{e1}
\end{equation}%
and%
\begin{equation}
\varepsilon _{2}\left( \omega ,k\right) =\min \left\{ \varepsilon _{F},\frac{%
M^{\ast }}{K_{\mu }^{2}}\left( \omega \,\omega _{R}+\sqrt{k^{2}\left( \omega
_{R}^{2}-K_{\mu }^{2}\right) }\right) \right\}  \label{e2}
\end{equation}%
are the limiting energies of a nucleon hole, coming from the kinematics of
the process. In the above $p_{F}$ and $\varepsilon _{F}$ stand for the Fermi
momenta and kinetic Fermi energies of neutrons and protons.

One can easily find that 
\begin{equation}
\func{Im}\Pi _{\Delta h}^{33}\left( 0,k\right) =0  \label{SimD}
\end{equation}%
because $\theta \left( \varepsilon _{2}\left( 0,k\right) -\varepsilon
_{1}\left( 0,k\right) \right) =0$.

The lowest-order contribuion to the pion self-energy can be obtained by
contraction of the above tensor with $K_{\mu }K_{\nu }$. The lowest-order
self-energy has the correct non-relativistic limit. Indeed, in the
non-relativistic limit only the $\func{Im}\Pi _{\Delta h}^{33}$ component
contributes. This gives 
\begin{eqnarray}
\func{Re}\Pi _{\Delta h}\left( K\right) &=&\frac{4f_{\pi N\Delta }^{2}}{%
9m_{\pi }^{2}}\frac{M^{\ast }M_{\Delta }^{\ast }+M^{\ast 2}}{2M^{\ast 2}}%
k^{2}F_{\Delta }^{\left( 1\right) }\left( \omega ,k;p_{F}\right) \,  \notag
\\
&&  \label{PiDN}
\end{eqnarray}%
To the lowest order in $p_{n}/M^{\ast }$ one can replace $\varepsilon \left(
p\right) \rightarrow M^{\ast }$, then the integral over the nucleon momentum
can be performed to give%
\begin{eqnarray}
F_{\Delta }^{\left( 1\right) }\left( \omega ,k;p_{F}\right) &\rightarrow &%
\frac{1}{4}\left( \Phi _{0}\left( \omega -\omega _{R},k;p_{F}\right) +\Phi
_{0}\left( -\omega -\omega _{R},k;p_{F}\right) \right)  \notag \\
&&  \label{nr}
\end{eqnarray}%
where $\Phi _{0}\left( \omega ,k;p_{F}\right) $ is the Migdal function (\ref%
{MigF}). By considering the static limit, $M_{\Delta }^{\ast },M^{\ast
}\rightarrow \infty $, of Eqs. (\ref{PiDN}) and (\ref{nr}) we find the
following expression%
\begin{equation}
\Pi _{\Delta h}^{\mathrm{st}\,}\left( K_{\mu }\right) =-\frac{8}{9}\frac{%
f_{\pi N\Delta }^{2}}{m_{\pi }^{2}}\frac{n\omega _{R}k^{2}}{\omega
_{R}^{2}-\omega ^{2}},  \label{pst}
\end{equation}%
in agreement with the well-established static form \cite{Eric}.

\end{document}